\documentclass[conference]{IEEEtran}
\usepackage{graphicx} 
\ifCLASSINFOpdf
\else
\fi

\usepackage{amsmath}

\begin{document}
%
\title{A Convolutional Framework for Mapping Imagined Auditory MEG into Listened Brain Responses}
%
%
%

\author{
    Maryam Maghsoudi$^{1}$, 
    Mohsen Rezaeizadeh$^{1}$, 
    Shihab Shamma$^{1}$ \\
    $^{1}$Electrical and Computer Engineering Department, University of Maryland, College Park, MD
}

%
%

\markboth{Journal of \LaTeX\ Class Files,~Vol.~14, No.~8, August~2015}%
{Shell \MakeLowercase{\textit{et al.}}: Bare Demo of IEEEtran.cls for IEEE Journals}
%



\maketitle

\begin{abstract}
Decoding imagined speech engages complex neural processes that are difficult to interpret due to uncertainty in timing and the limited availability of imagined-response datasets. In this study, we present a Magnetoencephalography (MEG) dataset collected from trained musicians as they imagined and listened to musical and poetic stimuli. We show that both imagined and perceived brain responses contain consistent, condition-specific information. Using a sliding-window ridge regression model, we first mapped imagined responses to listened responses at the single-subject level, but found limited generalization across subjects. At the group level, we developed an encoder–decoder convolutional neural network with a subject-specific calibration layer that produced stable and generalizable mappings. The CNN consistently outperformed the null model, yielding significantly higher correlations between predicted and true listened responses for nearly all held-out subjects. Our findings demonstrate that imagined neural activity can be transformed into perception-like responses, providing a foundation for future brain–computer interface applications involving imagined speech and music.
\end{abstract}


%
\IEEEpeerreviewmaketitle

%
%
%
%
\section{Introduction}
\IEEEPARstart{M}{ental} imagery refers to the voluntary process of internally generating sensory experiences in the absence of external stimuli, allowing individuals to hear sounds, imagine melodies, or rehearse speech by retrieving representations stored in memory \cite{kosslyn2001neural}. In contrast, perception involves processes driven by external sensory input. Understanding how the brain represents internally generated auditory content, and how these representations relate to actual perception, provides fundamental insights into neural computation, predictive processing \cite{Friston2005Theory, Bendixen2009Iheard, Hovsepyan2020Combining}, and mental simulation. Moreover, the ability to decode internal auditory imagery has significant potential for brain–computer interfaces (BCIs), particularly for communication in individuals unable to produce overt speech.

Decoding imagined auditory content from neural signals remains a challenging problem, in large part because imagery is more variable, less time-locked, and more weakly encoded than real sensory input. Imagined speech, for example, lacks reliable onset timing: participants may start imagining a word earlier or later on each trial, may prolong or compress imagined segments, and may replay internal speech with nonuniform tempo. This temporal uncertainty complicates alignment of neural responses. Previous work addressed this difficulty using dynamic time warping (DTW) to temporally align imagined speech to spoken speech \cite{martin2016word}. In the auditory domain, studies focusing on musically trained participants have shown that imagined melodies evoke neural patterns that recapitulate aspects of the acoustic structure, thereby offering a more tractable setting for analysis \cite{marion2021music, Herholz2012Neuronal, Kraemer2005Musical}. These findings demonstrate that internal auditory simulations share representational structure with real sensory responses, suggesting the feasibility of learning a mapping between them.

\begin{figure}[!t]
\centering
\includegraphics[width=3.5in]{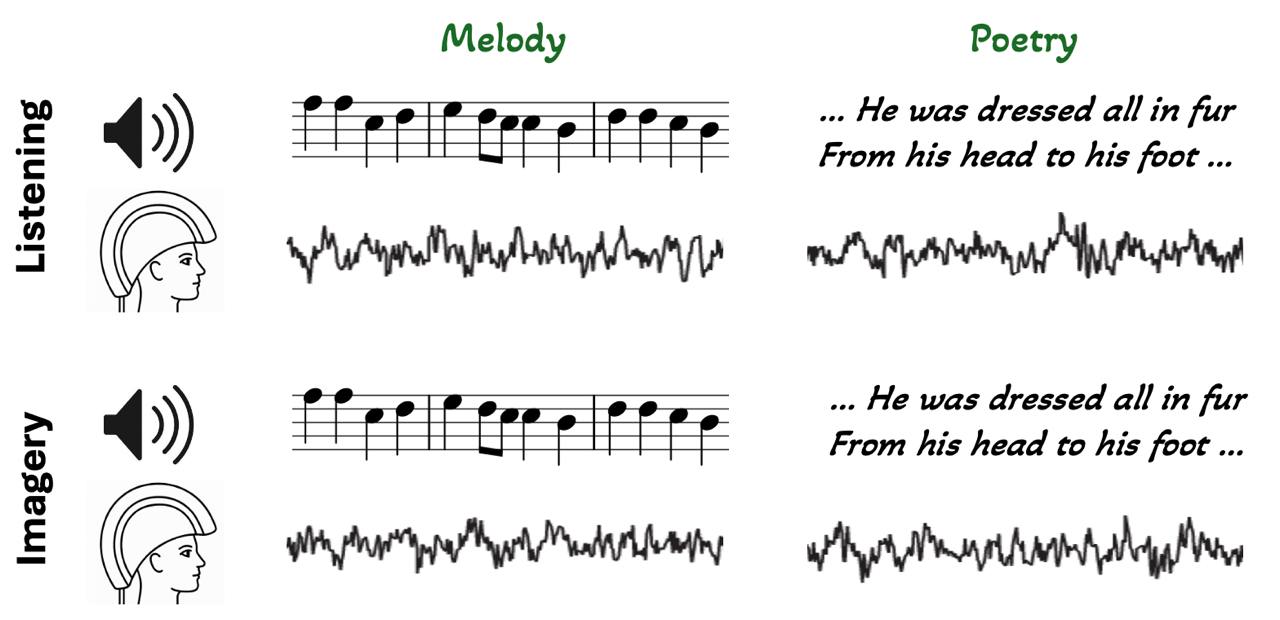}
\caption{MEG Experiment Paradigm. Some trials were listening condition (top) and the others were imagery condition (bottom). Participants listened to and imagined two melodies and two poem snippets.}
\label{exp}
\end{figure}

Recent years have seen substantial advances in decoding speech and acoustic features from noninvasive neural signals during listening. Deep learning models trained on MEG or ECoG can reconstruct phoneme sequences, spectrograms, or high-level linguistic features with increasing accuracy \cite{gwilliams2022neural, defossez2023decoding, yang2024mad, Akbari2019Towards, Angrick2019Speech}. However, most of these methods rely on time-locked, high-signal-to-noise responses during perception. Far fewer studies have attempted to decode imagined auditory content, and existing work is often limited by small datasets, reduced signal-to-noise ratio, and strong subject-specific variability. Moreover, most approaches focus on coarse classification of a small set of categories rather than reconstructing continuous imagined audio \cite{Dash2020Decoding}, \cite{Alharbi2024_decoding_imagined_speech}. 

\begin{figure*}[htbp]
\centering
\includegraphics[width=7in]{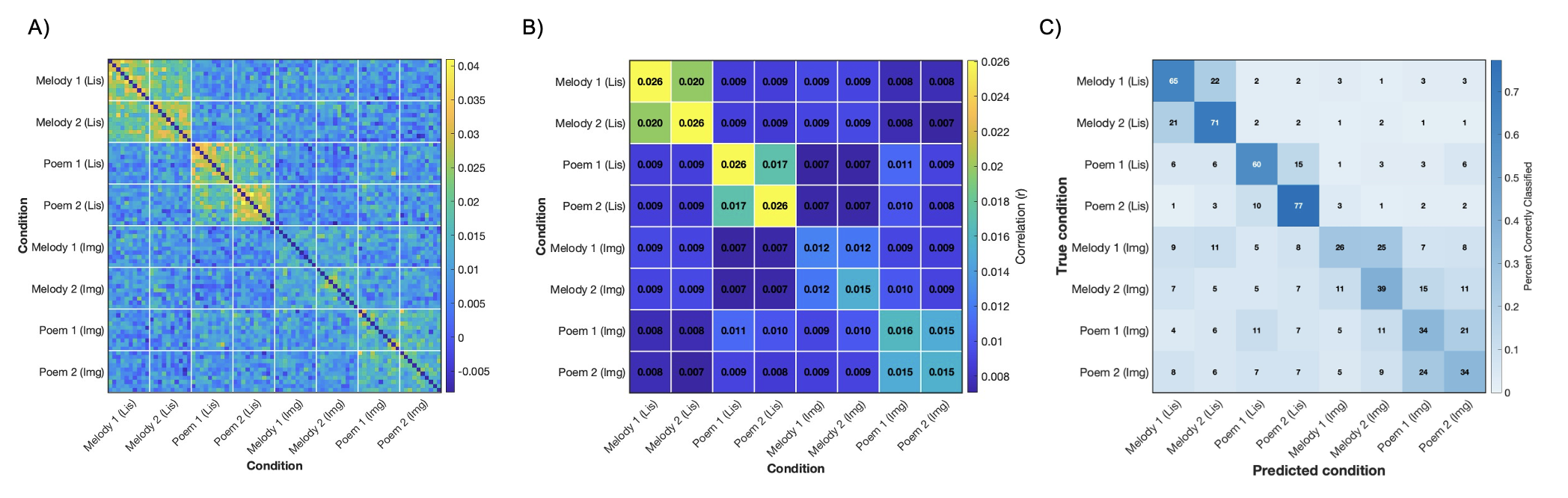}
\caption{A) Trial-by-trial representational similarity matrix computed across all MEG trials. Each 10×10 block corresponds to correlations within and between the eight experimental conditions.
B) Block-averaged representational similarity matrix (8×8), showing the mean correlation between conditions.
C) Confusion matrix for the correlation-based classifier. Each cell shows the percentage of trials assigned to each predicted condition, row-normalized so that values represent correctly classified (\%) for each true condition.}
\label{corrclassify}
\end{figure*}

A major open question in the field is whether imagined responses can be transformed into the "listened-response space," enabling speech-decoding models trained on listened data to operate on imagined input. If so, this would bypass the need for large labeled imagined-speech datasets, which are difficult to collect with consistent timing and ground-truth labels.

In this study, we address this question by investigating the relationship between imagined and listened brain responses to both music and spoken poetry. Using magnetoencephalography (MEG), we recorded 27-second trials in which participants either listened to or imagined auditory stimuli. Our goal was to learn a transformation that maps imagined responses into the space of listened responses, enabling us to predict the neural response that would have occurred had the subject actually heard the sound. Such a transformation provides a promising pathway toward decoding imagined auditory content using models trained exclusively on listened data.

We first evaluated the feasibility of this approach using regularized linear regression in a sliding-window framework. Linear models trained separately for each participant produced significant correlations between predicted and actual listened responses, demonstrating that imagined activity carry information predictive of sensory responses. However, these linear models failed to generalize across subjects, highlighting the need for a model capable of capturing shared temporal structure while adapting to individual spatial patterns.

To address this limitation, we developed a convolutional neural network (CNN) architecture incorporating an encoder–decoder backbone and a lightweight subject-specific calibration layer. The backbone learns a shared temporal mapping between imagined and listened responses across participants, while the calibration layer adjusts for subject-specific spatial variability. This design allows us to leverage multi-subject data to learn a generalizable representation while preserving flexibility needed for individual adaptation.

The main contributions of this work are as follows:
(i) We introduce a new MEG dataset containing both imagined and listened responses to both musical and spoken auditory stimuli.
(ii) We show that these MEG responses contain condition-specific structure: correlation-based representational analyses and classification reveal that imagined and listened trials are reliably distinguishable.
(iii) We evaluate a per-subject linear regression model that maps imagined responses to listened responses, demonstrating reliable within-subject prediction but limited cross-subject generalization.
(iv) To overcome this limitation, we develop a convolutional neural network with a subject-specific calibration layer that learns a shared cross-subject mapping while adapting to individual variability, substantially improving prediction performance on unseen subjects.
Together, these findings provide a framework for bridging imagination and perception in the brain, opening the door to noninvasive decoding of internally generated auditory experiences.

\section{Experimental Paradigm}
We recorded magnetoencephalography (MEG) data from 11 trained musicians (7 male), all of whom self-reported normal hearing and provided written informed consent before participation. All study procedures were reviewed and approved by the Institutional Review Board of the University of Maryland. Participants received monetary compensation for taking part in the study.

Each subject completed eight experimental conditions, arising from two melodic stimuli and two spoken-poem stimuli, each performed in both a listening and an imagery task. This yielded: melody 1–listened, melody 1–imagined, melody 2–listened, melody 2–imagined, poem 1–listened, poem 1–imagined, poem 2–listened, and poem 2–imagined. The melodic materials were adapted from a monophonic MIDI collection of Bach chorales (BWV 263 and BWV 354), while the spoken materials consisted of two different excerpts from the poem “A Visit from St. Nicholas” (Moore or Livingston, 1823). All audio files were processed with noise reduction and perceptually matched in loudness to ensure consistency across melodic and speech stimuli.

Before MEG recording, each participant completed an individualized training session designed to ensure reliable mental imagery performance. During this session, they practiced both listening and imagery versions of the task and were also asked to vocally reproduce the melodies and recite the poem excerpts so experimenters could verify accuracy and timing.

During the main experiment, all stimuli were 27 seconds in duration (Fig. \ref{exp}). Each participant performed 10 trials per stimulus for both listening and imagery, resulting in 40 listening and 40 imagery trials (80 trials total) presented in randomized order. A visual cue in the form of a clock-style metronome was displayed to help subjects maintain the original temporal structure of the stimulus.

MEG signals were collected using a whole-head KIT (Kanazawa Institute of Technology) system equipped with 157 axial gradiometers. Data were sampled at 1 kHz with an online 500-Hz low-pass filter and a 60-Hz notch filter. Recordings took place in a magnetically shielded room (VAC). Participants lay in a supine position to minimize movement-related artifacts. During listening trials, audio was delivered binaurally through Etymotic Research ER-2 insert earphones at a comfortable sound level of approximately 70 dB SPL (A moderate sound pressure level comparable to everyday conversational speech).

\section{Data Preprocessing}
MEG data were visually inspected for sensor malfunction, and any saturating or non-functioning channels were removed. Signals were bandpass filtered between 0.1--8~Hz using a zero-phase third-order Butterworth filter.

Denoising Source Separation (DSS)~\cite{de2008denoising} was then applied to improve trial-to-trial reliability; the first seven components, ranked by cross-trial consistency, were retained. The resulting data were z-scored per channel and subsequently downsampled to 100~Hz for analysis.

\section{Representational Similarity}
We began by evaluating whether the MEG responses contained sufficient condition-specific information in both listening and imagery trials. For each subject and each channel, we computed a trial-by-trial Pearson correlation matrix, comparing every trial with all others from the same subject and channel. These analyses revealed significantly higher correlations within conditions (e.g., Melody 2 Listening) than between conditions (e.g., Poem 1 Imagery vs. Poem 2 Listening), indicating that the neural signals carry distinct and consistent patterns across trials. We also observed that listened responses show high within-condition correlations, especially for the two melodies, whereas poem conditions show lower within-condition coherence, consistent with linguistic variability and weaker rhythmic entrainment (Fig. \ref{corrclassify}A,B).

To quantify condition-specific information, we performed a correlation-based classification. For each trial, we computed its correlation with all other trials and assigned it to the condition with the highest average correlation. Across subjects, correlation-based classification achieved a mean accuracy of \(50.7\%\) (SEM = \(4.1\%\)), significantly above the chance level of \(12.5\%\) (Wilcoxon signed-rank test, \(p \ll 0.001\)) (Fig. \ref{corrclassify}C).

\section{From Imagery to Listening}
Building on these representational findings, we next asked whether imagined MEG responses could be transformed into their corresponding listened responses. To test this, we evaluated several modeling approaches for learning a mapping from imagery to perception.

\subsection{Linear Modeling}
We began by implementing a regularized linear regression (ridge) model applied in a sliding-window framework to test whether short temporal segments of imagined responses can predict the corresponding listened responses. The mappings were fit separately for each subject, allowing us to assess subject-specific correspondence between imagination and perception. Each channel was z-scored across all trials and time samples to remove scale differences. Ridge regression was applied in temporal windows of $500$ ms with $100$ ms step. For each window the model learned a mapping $\hat{Y} = XW$ where $X$ is a matrix of imagined MEG responses and $W$ is estimated with $l2$ regularization. 

We evaluated prediction quality using Leave-One-Trial-Out cross-validation of the Pearson correlation between the predicted and true listened responses. To evaluate whether mapping performance exceeded chance, we created a null distribution by shuffling the labels of trials. The entire procedure was repeated using shuffled imagery data, producing a null distribution of mapping correlations which match the true analysis in structure. 

\begin{figure}[htbp]
\centering
\includegraphics[width=3.4in]{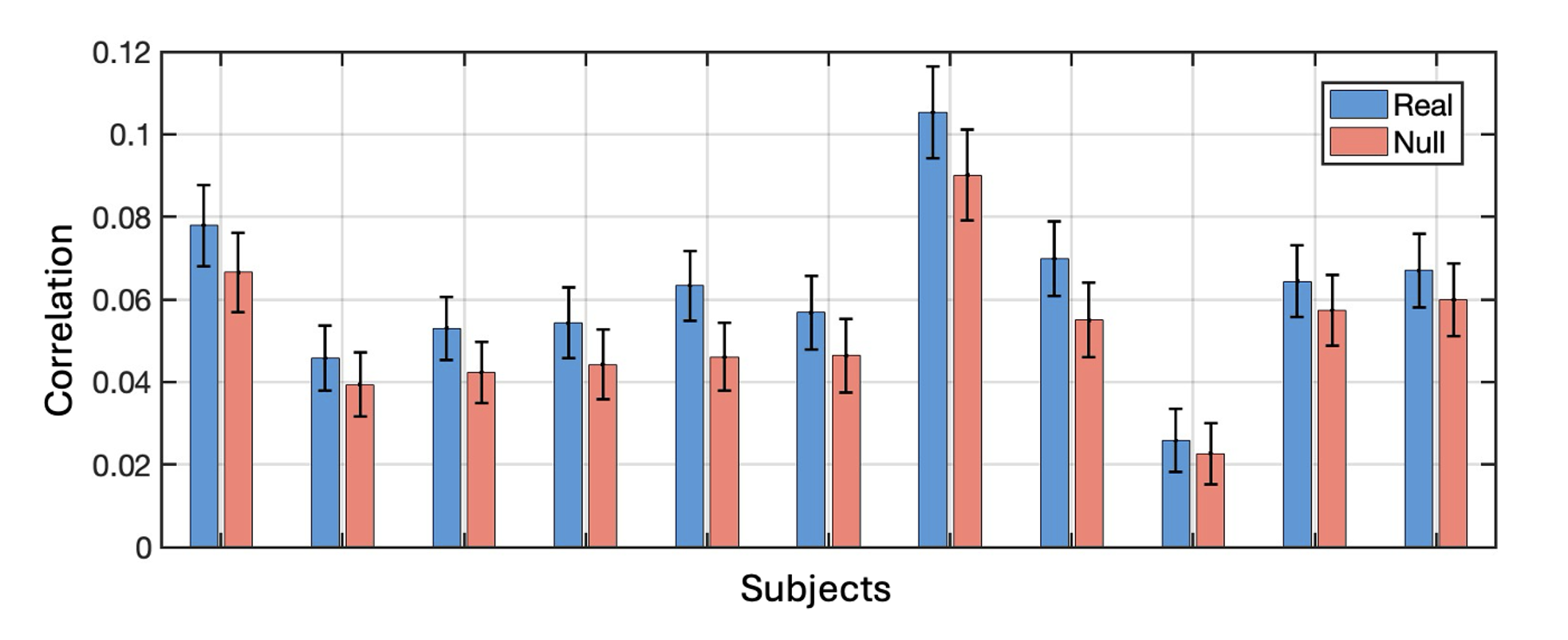}
\caption{Linear mapping performance across subjects. Mean sliding-window correlations between the predicted listened response and the ground-truth listened signal (Real) are shown alongside correlations from the shuffled null model (Null). Error bars indicate the standard error computed across windows and channels.}
\label{lin}
\end{figure}

Across subjects, the linear mapping achieved higher prediction accuracy for the real  pairs compared to the null model (Wilcoxon signed-rank test, \(p \ll 0.001\)) (Fig. \ref{lin}). 

\begin{figure}[htbp]
\centering
\includegraphics[width=3.49in]{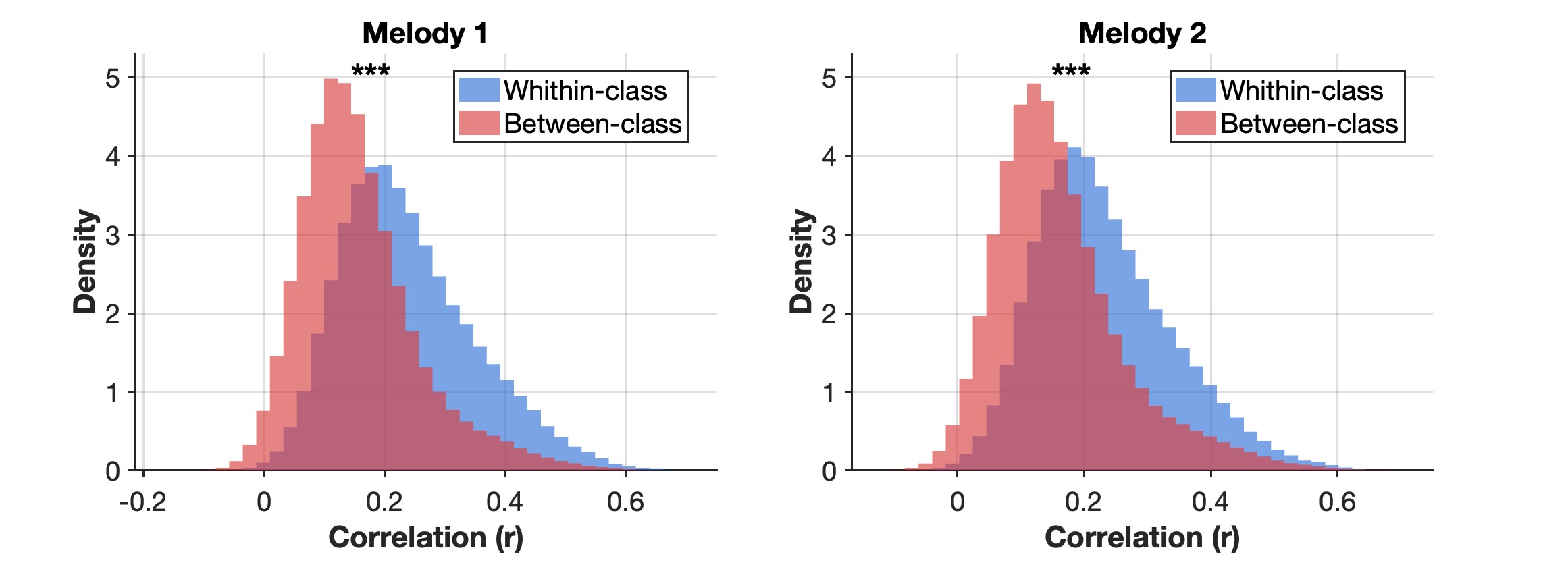}
\caption{Linear mapping.
Histogram of correlation coefficients between the predicted listened response and all listened trials. Blue: correlations with trials from the same class. Red: correlations with trials from all other classes.}
\label{predcorr}
\end{figure}

To further evaluate how well these mappings capture melodic structure, we examined whether the predicted listened responses correlated more strongly with trials from the same melody compared to trials from other melodies. As before, the model was trained in a leave-one-trial-out manner. For each held-out trial, we generated a predicted listened response and computed its correlation with all listened trials in the same class (within-class) and with trials from the other class (between-class). The resulting distributions are shown in Fig.~\ref{predcorr}. In both melodies, within-class correlations were significantly higher than between-class correlations (\(p \ll 0.001\)), indicating that the subject-specific linear mappings recover melody-specific structure in the predicted responses.

However, these linear mappings did not generalize across subjects, and training on all subjects and testing on a held-out subject yielded poor performance. This motivated the need for a more expressive deep learning architecture.

\subsection{Convolutional Neural Network}
\begin{figure}[htbp]
\centering
\includegraphics[width=3.4in]{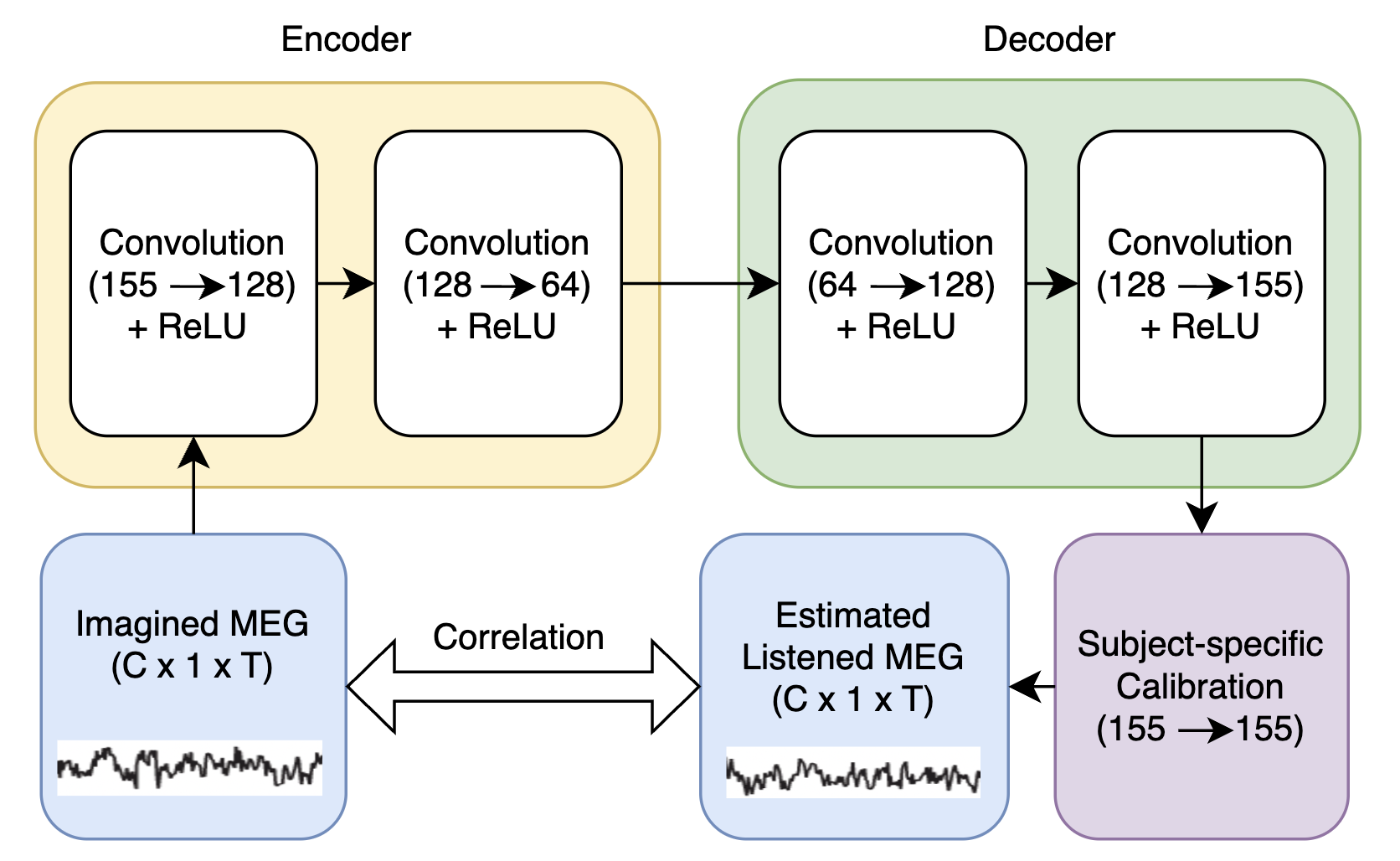}
\caption{Schematic of the encoder–decoder CNN and subject-specific calibration module used to map imagined to listened MEG responses. The encoder compresses the 155-channel input features, and the decoder reconstructs a 155-channel estimate of the listened response. A calibration layer adapts the shared model to each individual subject before final prediction.}
\label{cnn}
\end{figure}

The linear ridge regression model was able to predict listened MEG responses from imagined responses when trained separately for each subject. However, it did not generalize across subjects. To enable cross-subject generalization, we developed a CNN-based architecture to map imagined responses to their listened counterparts. The network consists of an encoder–decoder backbone with two one-dimensional temporal convolutional layers in each block, followed by a subject-specific calibration layer (Fig. \ref{cnn}). The model input is a tensor of shape $c \times 1 \times T$, where $C$ is the number of MEG channels and $T$ is the time length (27 seconds at 100 Hz). 

At each run, we held out one subject and constructed the training set by aggregating data from all other subjects. A train–validation split of 80–20 was applied. The encoder–decoder backbone was optimized using the Adam optimizer (learning rate $10^{-4}$). It was first trained on these data using $1D$ convolutions with kernel size 7 (approximately 70 ms) to capture local temporal context, ReLU activations, dropout rate 0.1, and early stopping based on validation loss. The training objective is a weighted combination of four complementary losses:
(i) Mean-squared error (MSE) to match amplitude structure,
(ii) Negative Pearson correlation loss to encourage correlation-based alignment independent of scale,
(iii) Temporal-difference loss to maintain smoothness and preserve local temporal derivatives,
(iv) spectral loss to encourage frequency-domain consistency between predicted and target signals:

\begin{align}
L_{\text{mse}} &= \sum_i \lVert \hat{y}_i - y_i \rVert^2, \label{eq:mse} 
\end{align}

\begin{align}
L_{\text{corr}} &= - \sum_i 
\left(\frac{ (x_i - \bar{x}_i)\,(y_i - \bar{y}_i) }
     { \lVert x_i - \bar{x}_i \rVert^2 \,\lVert y_i - \bar{y}_i \rVert^2 + \varepsilon }
\right), \label{eq:pearson}
\end{align}

\begin{align}
   L_{\text{temp}} &= \sum_i \lVert \Delta \hat{y}_i - \Delta y_i \rVert^2,
\qquad
\Delta y_i = y_{i,t+1} - y_{i,t}. \label{eq:temporal} 
\end{align}

\begin{align}
L_{\text{spec}} &= 
\sum_i \lVert |F(\hat{y}_i)| - |F(y_i)| \rVert^2, \label{eq:spectral}
\end{align}

\begin{align}
L_{\text{total}} &= 
L_{\text{mse}}
+ \alpha L_{\text{corr}}
+ \beta L_{\text{temp}}
+ \gamma L_{\text{spec}}. \label{eq:total}
\end{align}

The weighting coefficients $\alpha$, $\beta$, and $\gamma$ for $L_{\text{corr}}$, $L_{\text{temp}}$, and $L_{\text{spec}}$ were determined through an ablation study.

After backbone training, the calibration layer was trained on a small subset of the held-out subject’s data using the same objective. The calibration layer provides subject-specific adaptation while keeping the temporal mapping shared across the population. It is implemented as a single $1D$ convolution with kernel size 1, which performs a linear mixing of MEG channels. Conceptually, this operation reweights and recombines channels to account for differences in head position, anatomy, sensor alignment, and channel-specific signal characteristics across subjects. Because the backbone learns a generic temporal transformation from imagined to listened responses, the calibration layer only needs to adjust the spatial projection of the features for each new subject. This allows the model to preserve cross-subject generalization while remaining flexible enough to accommodate individual variability. 

\begin{figure}[htbp]
\centering
\includegraphics[width=3.49in]{cnn2222.jpg}
\caption{Fisher z-transformed correlations for the CNN on training and validation data under the true and null models. Each point represents a subject, and paired lines show within-subject differences.}
\label{cnn_res2}
\end{figure}

To establish a baseline, a null model was constructed by shuffling trial order and channel assignments. The backbone and calibration layer were retrained on the shuffled data using the same pipeline. Training and validation performance for subjects included in the training set are shown in Fig.~\ref{cnn_res2}. Fisher z-transformed correlations are reported for both the true and null models, with paired lines indicating within-subject differences. These results demonstrate that, for subjects the model has seen during training, the backbone learns a stable and meaningful mapping from imagined to listened responses and generalizes to unseen temporal segments of the same seen subjects.

\begin{figure}[htbp]
\centering
\includegraphics[width=3.4in]{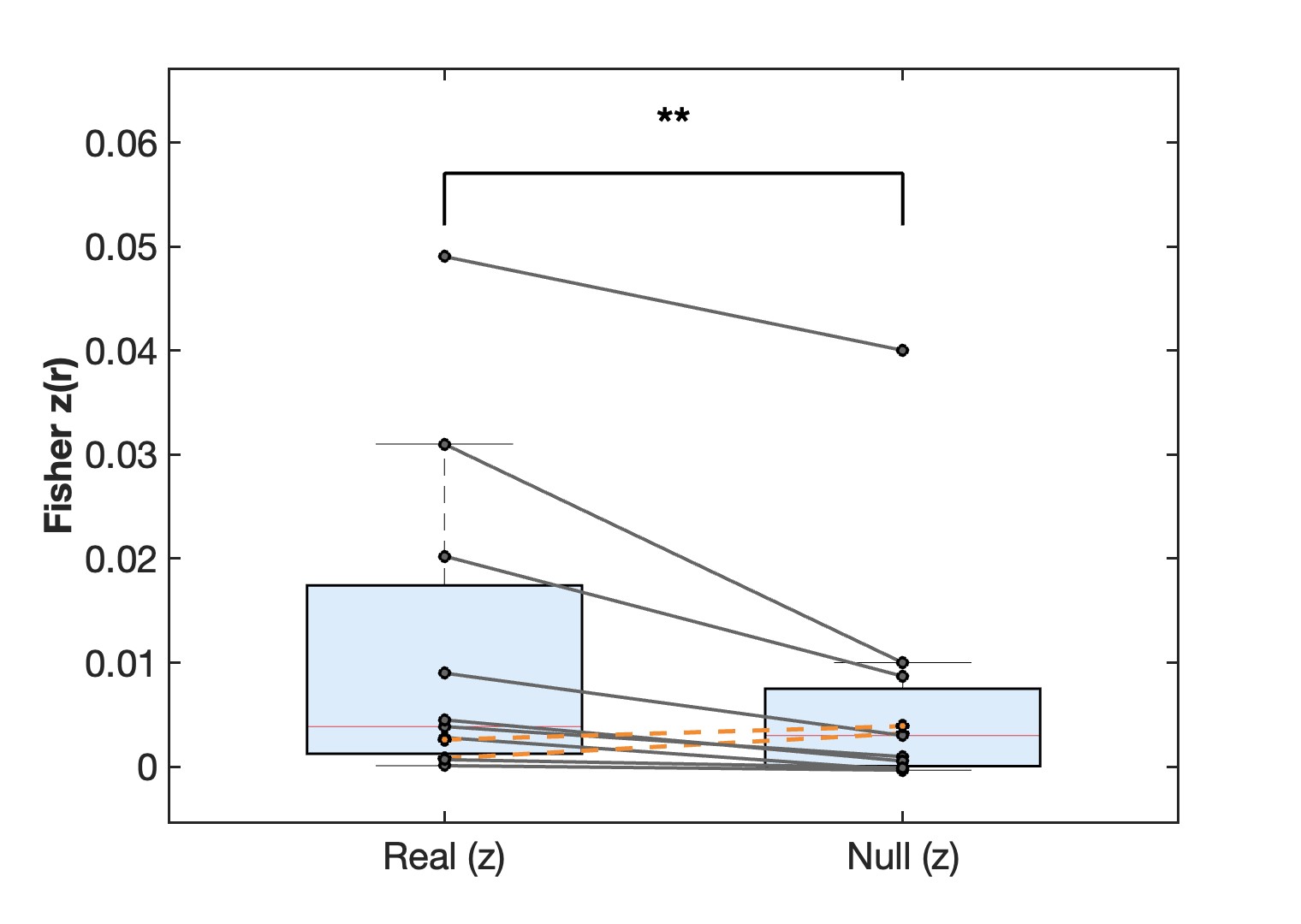}
\caption{Test-set performance of the CNN model compared against the null model. Fisher z-transformed correlations between predicted and true listened MEG responses are shown for each subject, with paired lines linking real and null values.}
\label{cnn_res1}
\end{figure}

To evaluate generalizability across subjects, the calibrated model was then evaluated on unseen data from the held-out subject by computing the Pearson correlation between predicted and true listened responses. This leave-one-subject-out procedure was repeated for all subjects. For each held-out subject, this produced a matched pair of correlation values for the true and null models. Fisher-transformed correlations were compared at the group level to stabilize variance. Except for two subjects, the true model exceeded the null model, with mean $r_{true} = 0.0113$, mean $r_{null} = 0.0063$, and Wilcoxon signed-rank test $p \ll 0.01$ (Fig. \ref{cnn_res1}). 

Taken together, these findings indicate that the CNN captures meaningful imagery–listening structure beyond chance and is able to transfer the learned mapping to previously unseen subjects

\section{Conclusion}
We investigated whether imagined auditory MEG responses can be transformed into their listened counterparts. Using a new dataset of imagined and perceived music and speech, we showed that the responses contain reliable condition-specific structure. We showed that a per-subject linear regression model could map imagined to listened responses, but did not generalize across individuals. To address this, we introduced a CNN with a subject-specific calibration layer, which learned a shared cross-subject mapping and significantly outperformed the null model. These results demonstrate that imagined responses preserve enough structure to predict corresponding listened activity, providing a foundation for future work on reconstructing imagined auditory content.

\section{Acknowledgement}
The authors acknowledge partial funding to SS for this work from AFOSR (FA9550-19-1-0408) and ONR (MURI: Learning from Soundscapes).

\ifCLASSOPTIONcaptionsoff
  \newpage
\fi



%
\bibliographystyle{IEEEtran}
\bibliography{references}

%

\begin{IEEEbiography}{Michael Shell}
Biography text here.
\end{IEEEbiography}

\begin{IEEEbiographynophoto}{John Doe}
Biography text here.
\end{IEEEbiographynophoto}


\begin{IEEEbiographynophoto}{Jane Doe}
Biography text here.
\end{IEEEbiographynophoto}




\end{document}